\documentclass[twocolumn,nofootinbib]{revtex4-1}
\usepackage{epsfig}

              \newcommand{\rf}[1]{(\ref{#1})}

\def\bfone{\relax{\rm 1\kern-.35em 1}}

\newcommand{\be}{\begin{equation}}
\newcommand{\ee}{\end{equation}}
\newcommand{\ben}{\begin{displaymath}}
\newcommand{\een}{\end{displaymath}}
\newcommand{\bea}{\begin{eqnarray}}
\newcommand{\eea}{\end{eqnarray}}

\newcommand{\bean}{\begin{eqnarray*}}
\newcommand{\eean}{\end{eqnarray*}}

\newcommand{\vp}{\varphi}

\def\K{K{\"a}hler}

\begin{document}

%\begin{center}
\title{\Large{Planck, LHC, and $\alpha$-attractors}}

\author{Renata Kallosh and Andrei Linde}

\affiliation{Department of Physics and SITP, Stanford University, \\ 
Stanford, California 94305 USA, kallosh@stanford.edu, alinde@stanford.edu}

\begin{abstract}
We describe a simple class of cosmological models called $\alpha$ attractors, which provide an excellent fit to the latest Planck data. These theories are most naturally formulated in the context of supergravity with logarithmic \K\ potentials. We develop generalized versions of these models which can describe not only inflation but also dark energy and supersymmetry breaking.
\end{abstract}

\maketitle

\smallskip

%\tableofcontents

\section{Introduction}\label{intro}

The results obtained by WMAP and Planck attracted growing attention to a mysterious fact that several different cosmological models proposed many years ago lead to almost exactly coinciding predictions, providing the best fit to most of the presently available observational data. This includes the Starobinsky model  \cite{Starobinsky:1980te}, the first model of chaotic inflation in supergravity proposed more than 30 years ago by Goncharov and Linde (GL model) \cite{Goncharov:1983mw,Linde:2014hfa}, and Higgs inflation \cite{Salopek:1988qh,Fakir:1990eg}. During the two years since the Planck 2013 data release, several broad classes of such models have been found and implemented in the context of supergravity and superconformal theory. We called them the cosmological attractors       \cite{Kallosh:2013hoa,Ferrara:2013rsa,Kallosh:2013yoa,Cecotti:2014ipa,Kallosh:2013tua,Galante:2014ifa,Kallosh:2015zsa}.  

\begin{figure}[ht!]
\vspace{-0.2cm}
\centering
{\hspace{5mm}
\includegraphics[scale=.4]{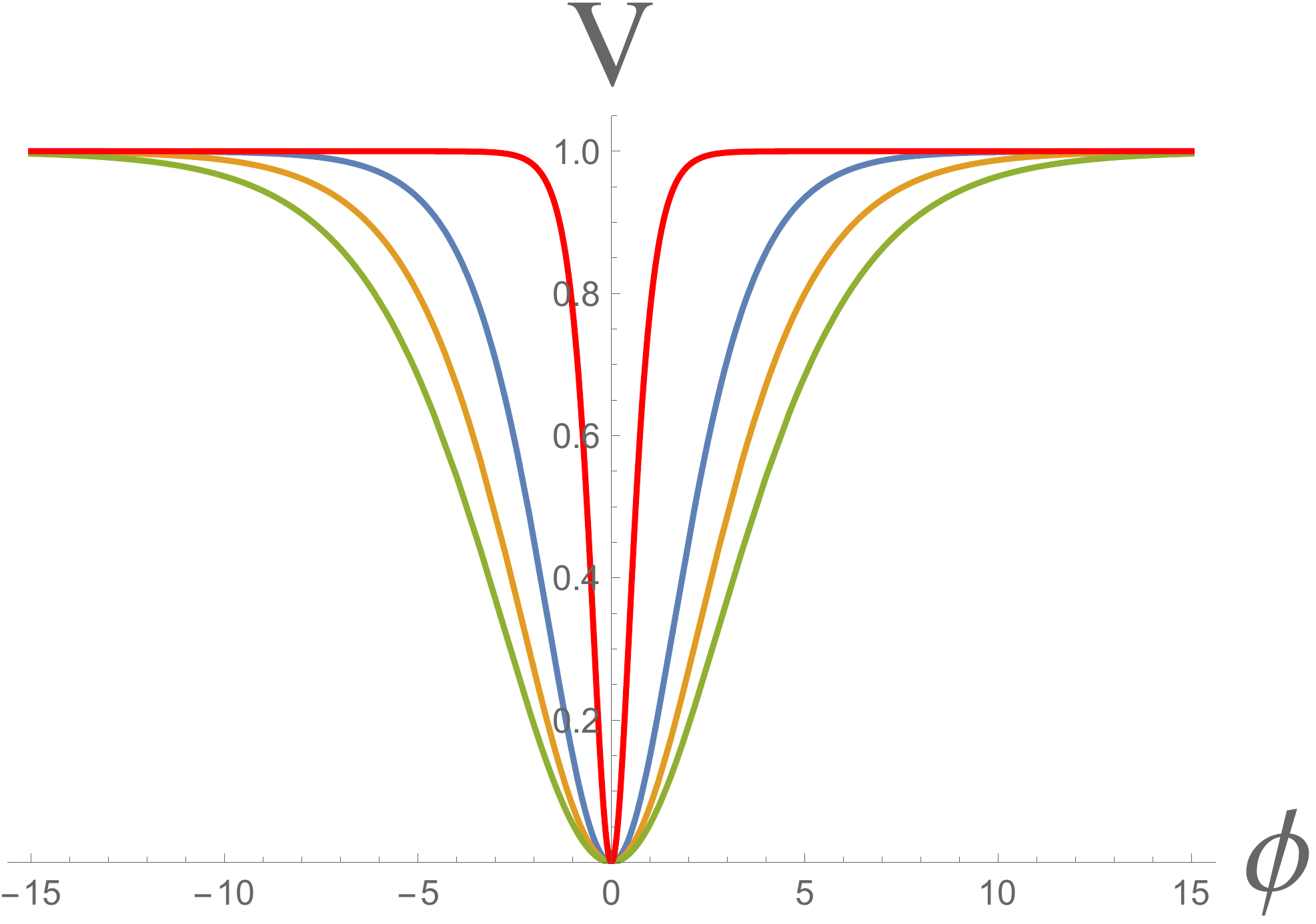}
\label{figT}}~~~~~
\caption{\footnotesize Blue, brown and green lines show the potentials of the T-models $\alpha \, \mu^2 \, \tanh^2{\varphi\over\sqrt {6 \alpha}}$ for $\alpha = 1, 2, 3$ correspondingly.  For comparison, the red line in the center shows the potential of the GL model  \cite{Goncharov:1983mw}, which is an $\alpha$ attractor with $\alpha = 1/9$. The potential is shown in units of $\alpha \mu^{2}$, the field is shown in Planck units. Smaller $\alpha$ correspond to more narrow minima of the potentials.}
\end{figure}

One of the most general classes of such models are $\alpha$-attractors  \cite{Kallosh:2013hoa,Ferrara:2013rsa,Kallosh:2013yoa}. 
The reason why these models can be interesting for cosmology becomes apparent when one studies their simplest representative, the {\it T-models}  \cite{Kallosh:2013hoa,Kallosh:2013yoa}
\be
  \mathcal{L}_{\rm E} = \sqrt{-g} \left[ {1\over 2} R - {(\partial \phi)^2\over 2(1-\phi^{2}/(6\alpha))^{2}}  - {m^2\over 2} \phi^2  \right] \, , \label{old Lag-alpha}
\ee
where $\alpha$ can take any positive value. In the large $\alpha$ limit, this theory coincides with the simplest model of chaotic inflation with a quadratic potential. A canonically normalized field $\varphi$ in this theory is related to the original field $\phi$ as follows: $\phi = \sqrt{6\alpha} \tanh{\varphi\over\sqrt {6 \alpha}}$. In terms of the canonically normalized field $\varphi$, this theory has a potential shown in Fig. \ref{figT}:
\be
V_{T} = \alpha \, \mu^2\tanh^2{\varphi\over\sqrt {6 \alpha}} \ , 
\ee
 where $\mu^{2} = 3m^{2}$.
 \begin{figure}[ht!]
%\vspace*{-1mm}%\hspace{-5mm}
\begin{center}
\includegraphics[width=9.2cm]{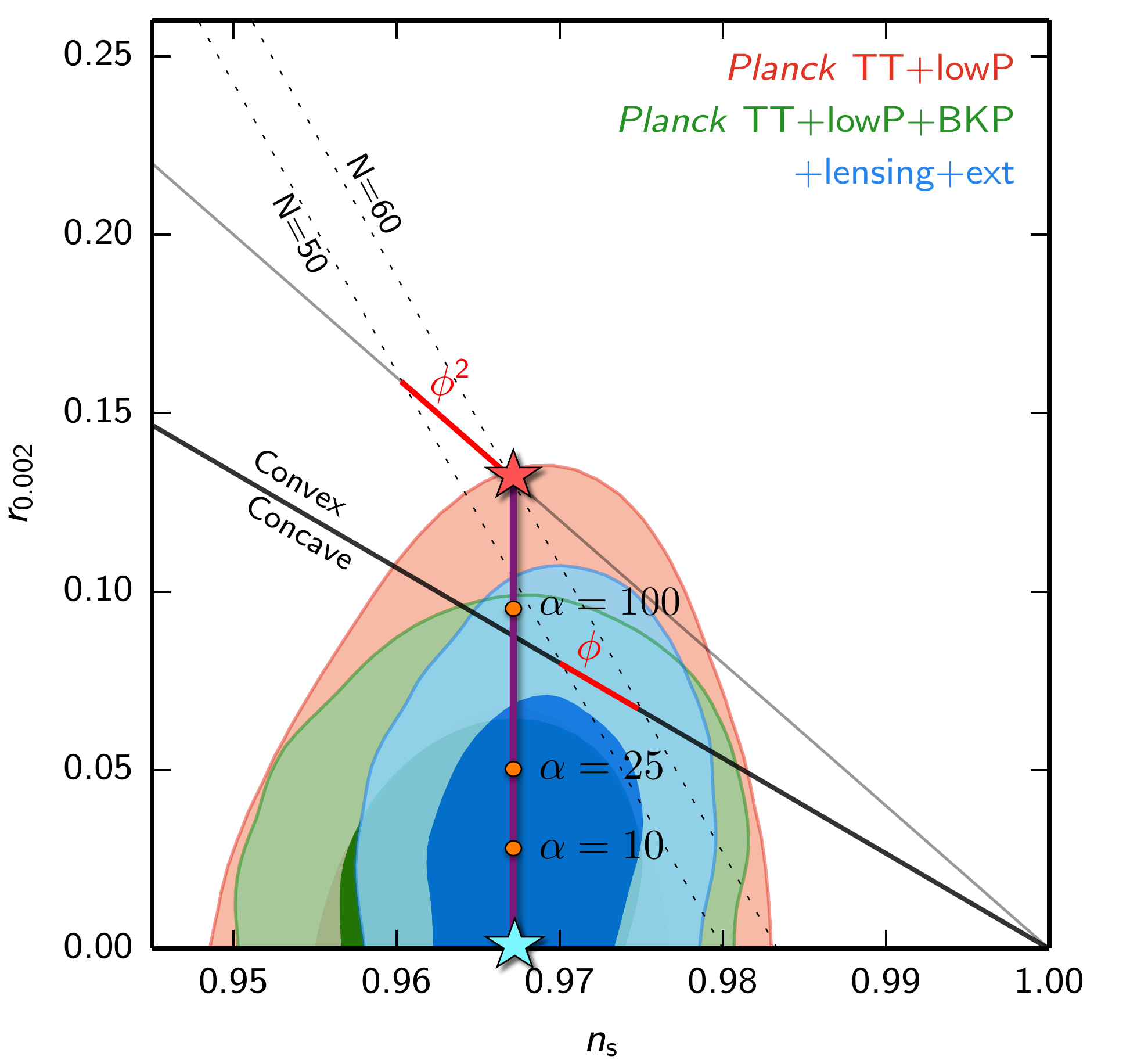}
\vspace*{-0.4cm}
\caption{\footnotesize Predictions of the simplest $\alpha$ attractor  T-models with the potential $V= \alpha\mu^{2} \tanh^2{\varphi\over\sqrt {6 \alpha}}$ cut through the most interesting part of the Planck 2015 plot for $n_{s}$ and $r$ \cite{Planck:2015xua}. They are shown as a purple { vertical line} starting at the predictions of the simplest quadratic model ${m^{2}\over 2}\phi^{2}$ for $\alpha > 10^{3}$ (red star), going down through $r \sim 0.05$ for $\alpha= 25$, $r \sim 0.027$ for $\alpha= 10$, through
the predictions of the Starobinsky model $r \sim 0.003$ for $\alpha= 1$, the predictions of the GL model $r \sim 0.0004$ for $\alpha= 1/9$, and continuing all the way down to $r \to 0$ for $\alpha \to 0$ (blue star). This line is shown for $N = 60$. }
\label{f1}
\end{center}
\vspace{-0.3cm}
\end{figure}

In the leading order in the inverse number of e-foldings $1/N$, for $\alpha \ll N$, the slow roll parameters $n_{s}$ and $r$ for T-models are
\be\label{nsr}
 1 -n_{s} = {2\over N}\, , \qquad r =  {12\alpha \over N^{2} } \ .
 \ee
For large $\alpha$, the prediction for $n_{s}$ practically does not change, but the growth of $r$ slows down: $r \approx  {12\alpha \over N(N+3\alpha/2) }$ \cite{Kallosh:2013yoa}.  The exact interpolating values of $n_{s}$ and $r$  for the theory $V = \tanh^2{\varphi\over\sqrt {6 \alpha}}$ are plotted in  Fig. \ref{f1} by a thick purple vertical line superimposed with the results for $n_s$ and $r$ from the Planck 2015 data release \cite{Planck:2015xua}. This line begins at the point corresponding to the predictions of the simplest quadratic model ${m^{2}\over 2}\phi^{2}$ for $\alpha > 10^{3}$ (red star), and then, for $\alpha\lesssim 40$, it enters the region most favored by the Planck data. For $\alpha = 1$, these models give the same prediction $r \sim 12/N^{2}$ as the Starobinsky model,  the Higgs inflation model \cite{Salopek:1988qh},  and the broad class of superconformal attractors \cite{Kallosh:2013hoa}. Then the same vertical line continues further down towards the prediction $r \sim 4/3N^{2}$ of the GL model \cite{Goncharov:1983mw,Linde:2014hfa} corresponding to $\alpha = 1/9$. Then it goes even further, all the way down to $r \to 0$ in the limit $\alpha \to 0$. The blue star in Fig. \ref{f1} covers simultaneously all above mentioned models with  $\alpha \lesssim 1$. 

One can show that not only $n_{s}$, but also the amplitude of scalar perturbations in this class of models in the large $N$ limit {\it does not depend on $\alpha$}; it depends only on $N$ and $\mu$. For $N = 60$, this amplitude matches the Planck 2015 normalization if $\mu \approx 10^{{-5}}$. 

Moreover, for sufficiently small $\alpha \lesssim O(1)$, the predictions of $\alpha$-attractors in the large $N$ limit almost do not depend on whether we take the potential   $\tanh^2{\varphi\over\sqrt {6 \alpha}}$, or use a general class of potentials $f^{2}(\tanh{\varphi\over\sqrt {6 \alpha}})$ for a rather broad set of choices of the functions $f$.  This stability of predictions, as well as their convergence  to one of the two attractor points shown in Fig. \ref{f1} by the red and blue stars, is the reason why we called these theories the cosmological attractors. The latest Planck 2015 result $n_{s} = 0.968\pm 0.006$ \cite{Planck:2015xua} almost exactly coincides with the prediction of the simplest T-models for $N = 60$. These properties of T-models are quite striking. Since their predictions can match any value of $r$  from $ 0.14$ to $0$, see Fig. \ref{f1}, these models may have lots of staying power.

As we already mentioned, the first model of this class was found more than 30 years ago  \cite{Goncharov:1983mw}. Later on, it was nearly forgotten because the plateau potentials have not been popular at that time. It took some time until the original version of the Starobinsky model \cite{Starobinsky:1980te} was reformulated as the theory $R+R^{2}$ and cast in the form with a very similar plateau potential for $\vp >0$. It took even longer until we learned several different ways to solve the problem of initial conditions for such models \cite{Linde:2004nz}. The general class of T-models and their attractor behavior were discovered only very recently.  From the point of view of the theory of fundamental interactions, it is interesting that these models naturally appear in the context of conformal and superconformal theories. In this context, the parameter $\alpha$ is related to the inverse curvature of the \K\ manifold \cite{Ferrara:2013rsa,Kallosh:2013yoa,Cecotti:2014ipa}.  
The attractor behavior resulting in stability of  predictions with respect to various deformations of potentials is a result of a nontrivial structure of the moduli space with a boundary \cite{Kallosh:2013hoa,Kallosh:2013yoa}. Equivalently, one can interpret the existence of attractors as a consequence of the existence of a pole in the kinetic term for the inflaton field in \rf{old Lag-alpha}   \cite{Galante:2014ifa}. 

Now that inflationary predictions of $\alpha$-attractors are well understood, one may wonder whether one can take a next step and generalize these models to achieve two additional goals. First of all, the potentials $V\sim \tanh^2{\varphi\over\sqrt {6 \alpha}}$ vanish in the minimum at $\vp =0$, but we would like to describe a universe with a tiny but non-zero vacuum energy $V_{0 }\sim 10^{-120}$. 
Secondly, many particle phenomenologists assume that we live in the world with weakly broken supersymmetry, with the gravitino mass $m_{3/2} \sim 10^{{-13}}-10^{{-15}}$ in Planck units. This assumption of the low value of SUSY breaking will  be tested at LHC during the next few years. 
However in the simplest supergravity versions of the T-models supersymmetry is unbroken,  $m_{3/2} =0$. 

This is not a real problem since the difference between $10^{-120}$,  $10^{{-13}}$ and $0$ is pretty small, so we are almost there already. One can always make a small remaining step by adding some new fields to the system, such as the Polonyi fields, to break SUSY and uplift the potential, see e.g. \cite{Dudas:2012wi}. However, this would force us to study a combined evolution of many moduli fields and strongly stabilize the Polonyi field to avoid the cosmological Polonyi field problem, which bothered cosmologists for more than 30 years.

An alternative solution is to utilize new possibilities offered by the recent cosmological constructions involving a nilpotent chiral multiplet which describes a Volkov-Akulov goldstino fermion \cite{Volkov:1973ix,rocek,Komargodski:2009rz} and has no fundamental scalars, see  \cite{Ferrara:2014kva,Kallosh:2014via,Dall'Agata:2014oka,Kallosh:2014hxa,Lahanas:2015jwa} for cosmological applications.

This possibility can be studied for T-models models using canonical \K\ potentials such as  $(\Phi-\bar\Phi)^{2}$.
However, even though it is possible to reproduce T-models in theories with such \K\ potentials, the main feature defining $\alpha$-attractors (a singular boundary of the moduli space) does not naturally emerge in this context. Therefore in this paper we will concentrate on logarithmic \K\ potentials, which more naturally appear in the context related to extended supergravity and string theory and naturally lead to attractor models. In this paper we will only briefly describe the main results of our investigation, leaving many details for a subsequent publication \cite{prep}.

\section{T-models  with unbroken SUSY and vanishing vacuum energy}

Here we will describe a supergravity realization of a T-model based on the theory of the field $\Phi$ coupled to the  field $S$ with the following \K\ potential:
\be\label{ka}
K= -3  \log \Big (1- Z\bar Z  + {\alpha-1\over 2} {(Z-\bar Z)^2\over 1- Z\bar Z}- {S\bar S\over 3}\Big )\, . 
\ee
The simplest T-model has a superpotential 
\be\label{cic}
 W= \sqrt \alpha\,  \mu S Z(1-Z^{2}) \ .
\ee
The field $S$ can be stabilized at $S=0$ by adding to the \K\, potential a term of the form $(S\bar S)^2$. Alternatively, one can take a nilpotent superfield $S$, satisfying $S^2(x, \theta)=0$ condition. In both cases 
 the potential of the field $z = {\rm Re \, Z}$ in this theory expressed in terms of a canonically normalized inflaton field $\vp$ is 
\be\label{nobreak}
V(z) = \alpha \, \mu^{2}z^{2} = \alpha \, \mu^{2} \tanh^{2} {\vp\over \sqrt {6\alpha}}\ , 
\ee
where $z = \tanh {\vp\over \sqrt {6\alpha}}$. Vacuum energy vanishes and supersymmetry is unbroken in the minimum of the potential with $S = 0$, $Z = 0$.    

A different supergravity embedding of the $\alpha$-attractor T model with an identical inflaton potential was given earlier in \cite{Kallosh:2013yoa}:  
\be\label{old}
K= -3\alpha  \log \Big (1- Z\bar Z - {S\bar S\over 3\alpha }\Big )  
\ee
and 
 \be\label{asuper}
 W=  \sqrt \alpha \,  \mu \, S Z\, (1-Z^{2})^{(3\alpha-1)/2}  \ .
 \ee

In some cases, one should add terms such as  $S\bar S{(Z-\bar Z)^2\over 1- Z\bar Z}$ or $S\bar S{(Z-\bar Z)^2\over (1- Z\bar Z)^{2}}$ to the  \K\, potential  for stabilization of the imaginary component of the field $Z$.

%The results for $n_{s}$ and $r$ for these models agree beautifully with the latest Planck data for a broad range of values of the parameter $\alpha \lesssim 40$: see  Fig. \ref{f1}. The amplitude of perturbations of the metric is proportional to $V^{3/2}/V' \approx {N \mu \sqrt{2\over 3}}= 49 \, \mu  $.  The Planck normalization \cite{Planck:2015xua} suggest that 
%$
%V^{3/2}/V' =  5.3 \times 10^{{-4}} 
%$
%which yields $\mu \approx  10^{{-5}}$, for $N = 60$.

At the minimum at $\vp =0$ in this model supersymmetry is unbroken, $D_ZW= D_SW=W=0$. This is fine if the field $S$ is stabilized at $S = 0$ in accordance with  \cite{Kallosh:2013yoa}.  However, if one would like to use a nilpotent field $S$, one should break SUSY at the minimum, which is what we were planning to do anyway. 

\section{T-models  with broken SUSY and non-vanishing vacuum energy}

If we want to use the same class of models, but describe simultaneously two other effects, SUSY breaking and non-vanishing cosmological constant, we should be prepared to pay the price. One can use the same \K\ potentials \rf{ka}, or \rf{old},
but one should modify the superpotential. Technically it means that $W$ must contain a term independent on $S$, in addition to the term linear in $S$, which was already present in inflationary models. Since we deal with $\alpha$-attractors, one can make some changes in the inflaton potential without altering the observational predictions. Therefore one can make the corresponding modification in several different ways. One option is to achieve a modification which reproduces exactly the same potential  $\alpha \, \mu^{2} \tanh^{2} {\vp\over \sqrt {6\alpha}}$ in the limit of small SUSY breaking. Yet another possibility is to allow modifications of the potential which do not change the observational predictions.

\subsection{Preserving the inflaton potential}\label{logkahl}

In order to reproduce the potential $\alpha\, \mu^{2} \tanh^{2} {\vp\over \sqrt {6\alpha}}$ (up to term with susy breaking in de Sitter vacua),  one can take
 \be\label{newsup}
 W=  \,   {\sqrt \alpha\mu\over 2} X^{3\over 2}  \Big (X^{3\sqrt \alpha\over 2} - c\, X^{-{3\sqrt \alpha\over 2}}\Big )  \Big (S+{1\over b}\Big )\, .
 \ee
 Here $X\equiv  1- Z^2$, and $c= 1- {2M\over \sqrt \alpha\, \mu }$.  For the special case $\alpha = 1$,  the superpotential of our model takes a  simple form
\be
 W= \Big (  \, {\mu\over 2} (X^3 -1)+M \Big )  \Big (S+{1\over b}\Big ) \ .
 \ee

A way to derive these expressions will be explained in a subsequent more detailed publication \cite{prep}; here we only present our main results. 
Leaving only the leading terms in the expansion in small parameters $M$ and $b^2-3$, one finds the inflaton potential 
\be\label{uplifted}
V=  \alpha \, \mu^{2} \tanh^{2} {\vp\over \sqrt {6\alpha}}+  {M^{2} (1-3/b^{2}}) +... 
\ee
The last term provides the required uplifting to a dS vacuum with a cosmological constant $V_{0} \sim 10^{{-120}}$. This term can be neglected during inflation. At the minimum at $\vp =0$,  supersymmetry is spontaneously broken,
\be
D_S W =  M\ , \quad  D_Z W = 0\ , \quad m_{3/2}=  {M\over b}\ ,
\label{dSpar}\ee
and vacuum energy is non-zero, 
\be
V_{0}=   M^{2} -3m^{2}_{3/2}=  M^{2} \left(1-{3\over b^{2}}\right) \ .
\label{dSpar2}\ee
Note that the uplifting is proportional to $M^{2}$, so dS uplifting is possible only because supersymmetry is spontaneously broken at the minimum.

Thus the main difference between the earlier model  \rf{cic} and the new model  \rf{newsup} is in the structure of the superpotential involving  two new parameters, SUSY breaking parameter $M$ and the parameter $b$ controlling the value of the cosmological constant $\Lambda = V_{0}= M^{2} \left(1-{3/ b^{2}}\right)$.  
In earlier models of  inflation in supergravity such as \rf{cic} the minimum of the potential typically was in a state with $\Lambda=0$ and unbroken supersymmetry with $DW=0$ in all directions in the moduli space. The origin of the universal positive contribution $M^{2}$ canceling the negative gravitino term $-3m^{2}_{3/2}$ in the potential \rf{dSpar2},  is the consequence of the presence in our models of the purely fermionic goldstino multiplet  \cite{Volkov:1973ix} of the Volkov-Akulov type.

Alternatively, one may consider models with the \K\ potential \rf{old}
and a slightly different superpotential,
 \be\label{WWW}
 W=    {\sqrt \alpha\mu\over 2} X^{3\alpha\over 2}  \Big (X^{3\sqrt \alpha\over 2} - c\, X^{-{3\sqrt \alpha\over 2}}\Big )  \Big (S+{1\over b}\Big )\, .
 \ee
Here, as before, $X\equiv  1- Z^2$, and $c= 1- {2M\over \sqrt \alpha\, \mu }$.  SUSY breaking is described by \rf{dSpar} and the potential is given by eq. \rf{dSpar2}. 

The inflaton potentials are the same in both cases, but in \rf{newsup} the stabilization of the inflationary trajectory occurs automatically (i.e. without adding stabilization terms such as $S\bar S{(Z-\bar Z)^2\over 1- Z\bar Z}$) for small $\alpha$, whereas in the models \rf{WWW} stabilization occurs automatically for large $\alpha$  \cite{prep}.

\subsection{Modifying the inflaton potential while preserving the cosmological predictions}

Now we will consider a model with a different superpotential  with \K\, in \rf{ka}:
\be
 W=  X\Big ( {1\over b} X+  S \Big )  (\sqrt \alpha \mu Z^2 + M)\ ,
\label{NewZ}\ee
with $X = 1-Z^{2}$. In this theory, just as in the model studied above, the magnitude of uplifting and SUSY breaking parameters are given by \rf{dSpar}, \rf{dSpar2}. Meanwhile the expression for the inflaton potential for $b^{2}\approx 3$ and $M\ll \mu$ is somewhat different:
\be\label{zz}
V(z) = {\mu^2 z^2\over 9} (4 - 16 z^2 + 3 (7 + 3\alpha)z^4 - 9 z^6) \ ,
\ee
where $z= \tanh {\vp\over \sqrt {6\alpha}}$. However, the height of the potential at the boundary of the moduli space at $z =1$ remains the same as in the theory \rf{nobreak}: $V(1) = \alpha\mu^{2}$. Therefore one can show that due to the magic of $\alpha$ attractors, the observational predictions of this model remains the same as in the simplest model \rf{nobreak}. As we will show in a separate publication, it happens not for all values of $\alpha$, but it does happen in the two most interesting cases, for $\alpha = O(1)$ and for $\alpha \gg 1$. An advantage of this version of the model is that the superpotential is represented by a rather simple function, and the \K\ potential does not require any stabilization terms.

The same inflaton potential \rf{zz} can be obtained also in the theory with the \K\ potential \rf{old} with a slightly modified superpotential:
\be
 W=  X^{3\alpha-1\over 2}\Big ( {1\over b} X+  S \Big )  (\alpha \mu Z^2 + M)\ ,
\label{NewZ1}\ee
with the dS uplifting given by \rf{dSpar2}.

 \subsection{Models with  logarithmic  superpotentials}

One can obtain the T-model with the simplest potential \rf{nobreak} using an alternative approach, which is based on a \K\ potential containing both the logarithmic and the power law part:
\be
K= -3  \log \Big (1- Z\bar Z  + {\alpha-1\over 2} {(Z-\bar Z)^2\over 1- Z\bar Z}\Big ) +{S\bar S}   \ .
\label{ANewSLog}\ee
This is achieved by using the superpotential which has a $\ln (1-Z^{2})$ contribution
\be\label{logsup}
W=    X^{3\over 2}  \Big ({\sqrt {3\alpha}\over 2}\,  \mu \,  { b \,}   \ln X +M \Big) \Big ({1\over b} +  S\Big ) \ ,
\ee
where  $X = 1-Z^{2}$. The potential at the minimum at $Z=0$,  and the parameters describing SUSY breaking,  as in all of our models, are given by \rf{dSpar}.
Thus when $b= \sqrt 3$ the cosmological constant vanishes. The axion field $Z-\bar Z$ is heavy during inflation where the potential depends on $z= \tanh {\vp\over \sqrt {6\alpha}}$ and the inflaton potential at $b^2=  3$ (i.e. ignoring the uplifting) is described by equation  \rf{nobreak}. 
Yet another way to reach the same goal is to use a different \K\ potential,
\be
K= -3\alpha  \log \Big (1- Z\bar Z  \Big ) +{S\bar S}   \ 
\ee
and a slightly different superpotential,
\be\label{logsup2}
W=    X^{3\alpha\over 2}  \Big ({\sqrt {3}\, \alpha\over 2}\,  \mu \,  { b \,}   \ln X +M \Big) \Big ({1\over b} +  S\Big ) \ .
\ee
In both cases, the inflaton potential is simple, as in Section \ref{logkahl}, but the price for this simplicity is the ``hybrid'' form of the \K\ potential involving both canonical and logarithmic terms, and the presence of an unusual logarithmic term in the superpotential.

\section{E-models}

\begin{figure}[htb]
\vspace*{-2mm}%\hspace{-3mm}
\begin{center}
\includegraphics[width=9.1cm]{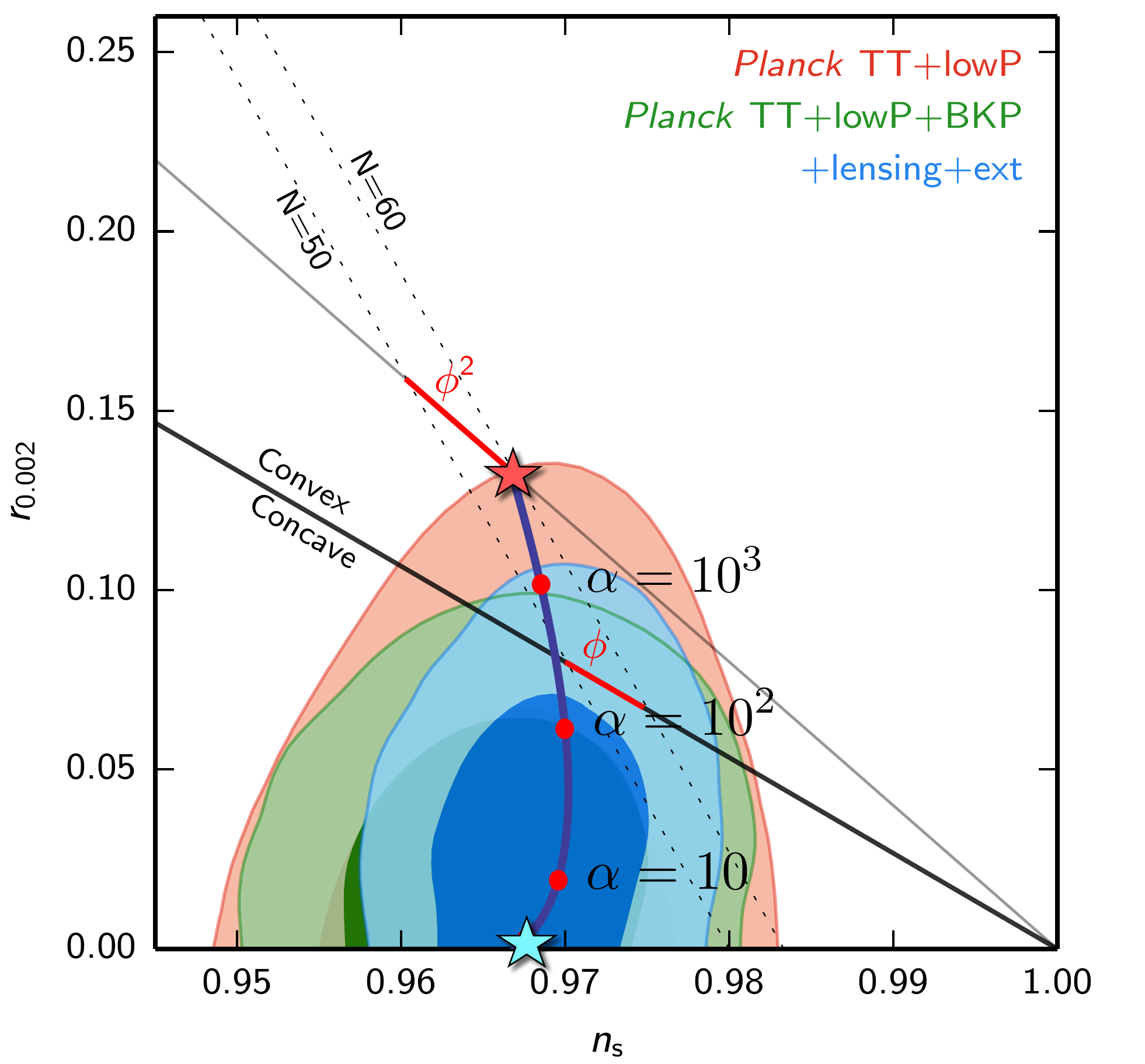}
\vspace*{-0.4cm}
\caption{\footnotesize Predictions of  E-models with $V\sim (1-e^{-{\sqrt {2\over 3 \alpha}}\varphi})^{2}$. They are shown as a thick blue curve starting at the predictions of the simplest quadratic model ${m^{2}\over 2}\phi^{2}$ for $\alpha > 10^{3}$, going down through the predictions of the Starobinsky model $r \sim 0.003$ for $\alpha= 1$, the predictions of the E-model generalization of the GL model \rf{appGE} $r \sim 0.0004$ for $\alpha= 1/9$ and continuing all the way down to $r \to 0$ for $\alpha \to 0$. This line is shown for $N = 60$. The red circles, from bottom up correspond to $\alpha = 10$, $\alpha = 10^{2}$ and $\alpha = 10^{3}$.}
\label{f2}
\end{center}
\vspace{-0.5cm}
\end{figure}

Yet another class of $\alpha$-attractors is equally interesting. Instead of explicit dependence on $\tanh^2{\varphi\over\sqrt {6 \alpha}}$, the potential of such models depends on $e^{-{\sqrt {2\over 3 \alpha}}\varphi}$. Therefore in this paper we will call them  E-models. The potential in the simplest class of such models is given by
\be\label{EE}
V_{E} = \alpha\mu^{2} \left(1-e^{-{\sqrt {2\over 3 \alpha}}\varphi}\right)^{2} \ .
\ee
For $\alpha = 1$, this potential coincides with the potential of the Starobinsky model. A consistent implementation of E-models in supergravity for general $\alpha$ was found in \cite{Ferrara:2013rsa,Kallosh:2013yoa}. Predictions of these models are shown in Fig. \ref{f2}.

\begin{figure}[ht!]
\centering
{
\includegraphics[scale=.41]{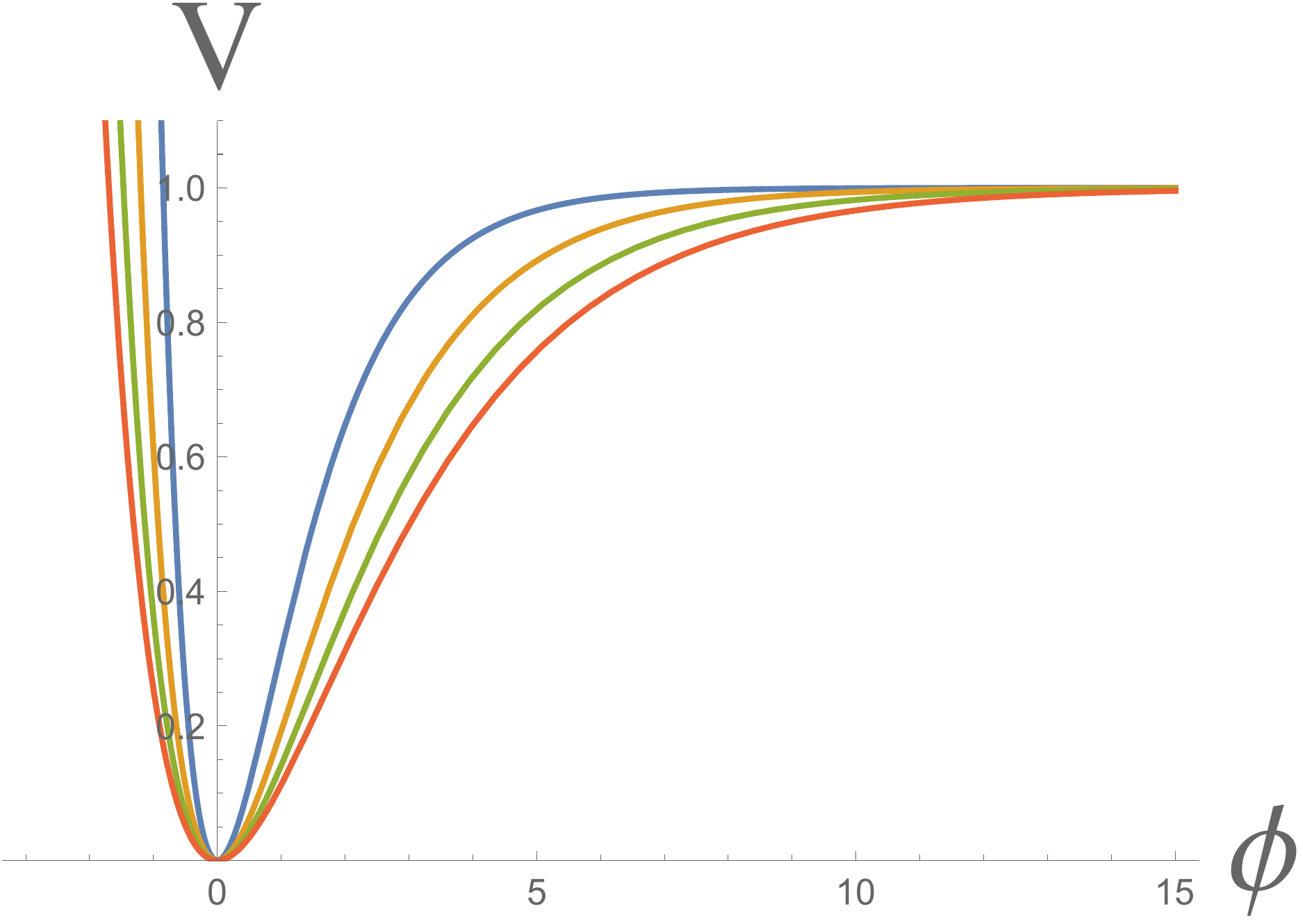}
\label{fig:newlongb}}
\caption{\footnotesize E-model potential  $\alpha\mu^{2}(1-e^{-{\sqrt {2\over 3 \alpha}}\varphi})^{2}$ in units of $\alpha\mu^{2} = 1$ for $\alpha = 1, 2, 3, 4$. Smaller $\alpha$ correspond to more narrow minima of the potentials.}
%\vspace{-0.2cm}
\end{figure}

As we see, predictions of T-models and E-models are similar, but not identical. The difference follows from the different shape of the inflationary potentials, see Fig.~\ref{figT} and Fig.~\ref{fig:newlongb}.

The simplest E-model (for $b^2=3$ with $V=0$ at the minimum, for simplicity) has the \K\ potential \cite{Cecotti:2014ipa}
\be
K= -3  \log \Big (T+\bar T  + {\alpha-1\over 2} {(T-\bar T)^2\over T+\bar T}\Big ) +{S\bar S}\, , 
\label{ANewSLog2}\ee
and superpotential
\be\label{Tamv}
W= \sqrt{2T\over 3}   (1 +  \sqrt 3 S)\Big ( 3 \,\mu\, \alpha\, (T\ln T - T +1) +  2 MT  \Big  ) .
\ee
The inflaton potential in terms of the canonically normalized field $\vp$ is given by \rf{EE}.
For the particular case $\alpha=1$, this model was presented in \cite{Lahanas:2015jwa}; its potential coincides with the potential of the Starobinsky model. Our generalization of this model allows to describe all points along the blue line in Fig. \ref{f2}. As one can see from this Figure, the predictions of these models are in very good agreement with the Planck 2015 data for $\alpha\lesssim 10^{2}$.

Alternatively, one can use the purely logarithmic \K\ potential \rf{ka} and the superpotential
\be
W = \bigl(\sqrt \alpha \mu Z^{2} +M\bigr)\Bigl({1\over \sqrt 3}(1+2Z)(1-Z)^{2}+ S(1-Z^{2})\Bigr).
\ee
The potentials of these models are more complicated than \rf{EE}, but they lead to the same observational predictions as the simplest E-models \rf{EE} for $\alpha = O(1)$ and for $\alpha \gg 1$. An advantage of such models is the absence of a rather unusual logarithmic term in the superpotential \rf{Tamv}.

\section{A special case: Goncharov-Linde model with $\alpha = 1/9$}

We will conclude this paper with a discussion of the GL model  \cite{Goncharov:1983mw,Linde:2014hfa}. From the point of view of the general classification outlined above, this model represents a single-field $\alpha$-attractor with  $\alpha = 1/9$. Original formulation of this model was based on the theory with
\be\label{shift}
K = -{1\over 2} (\Phi-\bar\Phi)^{2}\ ,\qquad  W = {\mu\over 9} \sinh{\sqrt{3}\Phi}\, \tanh{\sqrt{3}\Phi} \ .
\ee
This model has an interesting superconformal interpretation to be discussed in a  more detailed version of this paper. The inflaton potential in this model is given by
\be\label{potG}
V(\phi)= {\mu^2\over 27}  \Bigl(4 -  \tanh^{ 2}\sqrt{3\over 2} \vp\Bigr)\,  \tanh^{ 2}\sqrt{3\over 2} \vp\ \ .
\ee
It has a minimum at $\vp = 0$, where it vanishes, see Fig. \ref{figT}. At $\vp \gtrsim 1$, the potential coincides with 
\be\label{appG}
V(\vp) = {\mu^{2}\over 9} \left(1- {8\over 3}\, e^{-\sqrt 6 |\vp|}\right) \ ,
\ee 
up to exponentially small higher order corrections  $O(e^{-3\sqrt 6 |\vp|})$  \cite{Goncharov:1983mw}. These corrections can only lead to higher order corrections in $1/N$ to $n_{s}$ and $r$, where $N \sim 60$ is the number of e-foldings.   With our definition of the parameter $\mu$, the potential of this model matches the normalization of other $\alpha$ models in this paper, so that $V$ asymptotically approaches $\alpha\mu^{2} = \mu^{2}/9$. This model predicts $n_{s} = 1-{2\over N} \approx 0.967$ and $r = {4\over 3 N^{2}} \approx 4 \times 10^{{-4}}$ for $N \approx 60$, in excellent agreement with the Planck 2015 data.  It can describe not only inflation but also dark energy and SUSY breaking if one adds to it a nilpotent chiral multiplet with superpotential  \cite{Linde:2014hfa}.

Interestingly, the model with the GL inflaton potential \rf{potG} can be also obtained in the context of $\alpha$-attractors with a single-field logarithmic \K\ potential 
\be
K= -3  \log \Big (1- Z\bar Z  + {\alpha-1\over 2} {(Z-\bar Z)^2\over 1- Z\bar Z}\Big )  
\label{ANewSLogsingle}\ee
with $\alpha = 1/9$.
The superpotential in this representation of the GL model is particularly simple:
\be\label{GL}
 W =  {\mu\over 9}\, Z^{2}\, (1-Z^{2})  \ .
\ee
One can easily check that the inflaton potential of this model coincides with the potential of the original version of the GL model \rf{potG}, which is a T-model shown by the red line in Fig. \ref{figT}.

GL model allows various generalizations \cite{Linde:2014hfa}, which look especially simple in our approach. For example, if one one multiplies the superpotential \rf{GL} by $1+c Z$ with $|c|\ll 1$, the height of the plateau of the inflaton potential at $\vp > 0$ will be different from its height at $\vp < 0$. Furthermore, if one takes 
\be
W =\sqrt{2\over 3}  {\mu}\, Z^{2}\, (1-Z) \ ,
\ee
one finds the potential of an E-model
\be\label{appGE}
V(\vp) = {\mu^{2}\over 9} \left(1-  e^{-\sqrt 6 \vp}\right)\left(1- e^{-2\sqrt 6 \vp}\right) \ .
\ee 
Unlike the original GL model, the potential of this model depends not on $|\vp|$ but on $\vp$. This potential blows up at large negative  $\vp$,  has a minimum at $\vp = 0$, and approaches a flat plateau at large positive $\vp$, just as the family of potentials of E-models shown in Fig.~\ref{fig:newlongb}. This model has the same observational predictions as the original GL model.

 Note that the field $S$ is not required for the consistency of this family of models, which makes them most economical. However, the nilpotent field $S$ helps to break supersymmetry and uplift the minimum of the inflaton potential. This can be achieved, for example, by using the \K\ potential \rf{ka} and adding a simple $S$-dependent term to the GL superpotential \rf{GL}:
 \be\label{GLS}
 W =  {\mu\over 9}\, Z^{2}\, (1-Z^{2}) +M(S+1/b) \ .
\ee
This theory has the SUSY breaking parameters and the vacuum energy given by \rf{dSpar}, \rf{dSpar2}. Thus in this simple model one can simultaneously describe inflation, dark energy/cosmological constant, and SUSY breaking of a controllable magnitude.

%\newpage

\section{Discussion}

In this paper we discussed simplest models belonging to the general class of $\alpha$ attractors. These models lead to cosmological predictions providing excellent match to the latest cosmological data for a very broad range of $\alpha$. We described several different ways to implement such models in supergravity in a manner directly related to their attractor nature. We also developed  a set of $\alpha$-attractors describing not only inflation but also dark energy and supersymmetry breaking of a controllable strength. A more detailed description of our results will be given  in the subsequent publication \cite{prep}. 

The flexibility of the scale of supersymmetry breaking in these models may be important for considering an interplay between the cosmological data and the future data from LHC. The often made assumption of a small scale of SUSY breaking usually requires small reheating temperature, to avoid the cosmological gravitino problem. This constraint is removed if the gravitino mass is sufficiently large, in the range of $10^{2}$ TeV or above. In its turn, the reheating temperature affects the required number of e-foldings, and therefore the value of $n_{s}$. This effect is not large, but it may become noticeable with an increase of precision of the measurement of the cosmological parameters. In this way the results to be obtained at LHC may help us to optimize our choice of inflationary models based on supergravity.

\subsection*{Acknowledgements}
We are grateful to A. Achucarro, J.J. Carrasco, M. Galante, S. Dimopoulos, L. Hall, J. March-Russell, E. McDonough, M. Scalisi, U. Seljak and L. Senatore for important discussions. We are especially grateful to D. Roest for many enlightening comments and an ongoing collaboration on cosmological attractors. RK and AL are supported by the SITP and by the NSF Grant PHY-1316699. RK is also supported by the Templeton foundation grant `Quantum Gravity Frontiers,' and AL is supported by the Templeton foundation grant `Inflation, the Multiverse, and Holography.'

\end{document}